\documentclass[aps,prl,twocolumn,amsmath,amssymb,nofootinbib,showpacs,superscriptaddress]{revtex4}
\usepackage[english]{babel}
\usepackage{times}
\usepackage{latexsym}
\usepackage{graphics}
\usepackage{subfigure}
\usepackage{epsfig}
\usepackage{color}

\newcommand{\epsfboxmod}[1]{\epsfbox{#1.eps}}

\newcommand{\infig}[2]{\begin{center}
                                    \mbox{ \epsfxsize #1 \epsfboxmod{#2}}
                                      \vspace{-0.8cm}
                                    \end{center}}

\renewcommand{\i}{\textrm{i}}
\newcommand{\ie}{{i.e. }}

\newcommand{\ch}{\textrm{cosh}}
\newcommand{\sh}{\textrm{sinh}}

\newcommand{\Vh}{V_\textrm{ho}}
\newcommand{\Vopt}{V}
\newcommand{\Vr}{V_\textrm{R}}
\newcommand{\sigmar}{\sigma_\textrm{\tiny R}}

\newcommand{\LTF}{L_\textrm{TF}}
\newcommand{\xiini}{\xi_\textrm{in}}

\newcommand{\dist}{\mathcal{D}}

\newcommand{\lyap}{\gamma}
\newcommand{\loc}{L}

\newcommand{\kmob}{k_\textrm{m}}
\newcommand{\kc}{k_\textrm{c}}

\begin{document}

\title{Anderson Localization of Expanding Bose-Einstein Condensates in Random Potentials}

\author{L.~Sanchez-Palencia}
\affiliation{
Laboratoire Charles Fabry de l'Institut d'Optique,
CNRS and Univ. Paris-Sud,
Campus Polytechnique, 
RD 128, 
F-91127 Palaiseau cedex, France}
\homepage{http://www.atomoptic.fr}

\author{D.~Cl\'ement}
\affiliation{
Laboratoire Charles Fabry de l'Institut d'Optique,
CNRS and Univ. Paris-Sud,
Campus Polytechnique, 
RD 128, 
F-91127 Palaiseau cedex, France}

\author{P.~Lugan}
\affiliation{
Laboratoire Charles Fabry de l'Institut d'Optique,
CNRS and Univ. Paris-Sud,
Campus Polytechnique, 
RD 128, 
F-91127 Palaiseau cedex, France}

\author{P.~Bouyer}
\affiliation{
Laboratoire Charles Fabry de l'Institut d'Optique,
CNRS and Univ. Paris-Sud,
Campus Polytechnique, 
RD 128, 
F-91127 Palaiseau cedex, France}

\author{G.V.~Shlyapnikov}
\affiliation{
Laboratoire de Physique Th\'{e}orique et Mod\`{e}les Statistiques, 
Univ. Paris-Sud, F-91405 Orsay cedex, France}
\affiliation{\mbox{
Van der Waals-Zeeman Institute, Univ. Amsterdam, Valckenierstraat 65/67, 1018 XE Amsterdam, The Netherlands}}

\author{A.~Aspect}
\affiliation{
Laboratoire Charles Fabry de l'Institut d'Optique,
CNRS and Univ. Paris-Sud,
Campus Polytechnique, 
RD 128, 
F-91127 Palaiseau cedex, France}

\date{\today}

\begin{abstract}
We show that the expansion of an initially confined interacting 1D Bose-Einstein condensate
can exhibit Anderson localization in a weak random potential with correlation length $\sigmar$.
For speckle potentials the Fourier transform of the correlation function
vanishes for momenta $k>2/\sigmar$ so that the Lyapunov exponent vanishes
in the Born approximation for $k>1/\sigmar$.
Then, for the initial healing length 
of the condensate $\xiini>\sigmar$ the localization is exponential, 
and for $\xiini<\sigmar$ it changes to algebraic.
\end{abstract}

\pacs{05.30.Jp,03.75.Kk,03.75.Nt,05.60.Gg}

\maketitle

Disorder in quantum systems can have dramatic effects, 
such as strong Anderson localization (AL) of non-interacting particles in random media 
\cite{anderson1958}.
The main paradigm of AL is that
the suppression of transport is due to a destructive 
interference of particles (waves)
which multiply scatter from the modulations of a random potential.
AL is thus expected to occur when interferences
play a central role in the multiple scattering process
\cite{vantiggelen1999}.
In three dimensions, this requires the particle wavelength to be larger 
than the scattering
mean free path, $l$, as pointed out by Ioffe and Regel \cite{ioffe1960}.
One then finds a mobility edge at momentum $\kmob=1/l$,
below which AL can appear.
In one and two dimensions, all single-particle quantum states are predicted to be localized 
\cite{loconed,thouless1977,gang4}, 
although for certain types of disorder one has an 
effective mobility edge
in the Born approximation 
(see Ref.~\cite{izrailev1999} and below).
A crossover to the regime of AL has been observed in 
low dimensional conductors \cite{imry2002,gershenson1997}, and 
recently, evidences of AL have been obtained for light waves in bulk 
powders \cite{maret2006} and in 2D disordered
photonic lattices \cite{segev}.
The subtle question is whether and how the interaction between particles 
can cause delocalization and transport, and there is a long-standing discussion
of this issue for the case of electrons in solids \cite{basko2006}.

Ultracold atomic gases can shed new light on
these problems owing to an unprecedented control of interactions, 
a perfect isolation from a thermal bath, and the possibilities of designing
controlled random \cite{specklegrynberg,lye2005,clement2005,clement2006,fort2005}  
or quasi-random \cite{guidoni1997} potentials.
Of particular interest are the studies of
localization in Bose gases 
\cite{colddisorder2000s,lsp2005} and the interplay between interactions and disorder
in Bose and Fermi gases
\cite{interplay2000s,lsp2006}.
Localization of expanding Bose-Einstein condensates (BEC) in 
random potentials has been reported
in Refs.~\cite{clement2005,clement2006,fort2005}. However, 
this effect is {\it not} related to AL, 
but rather to the fragmentation of the core of the BEC, and 
to single reflections from large modulations of the
random potential in the tails~\cite{clement2005}.
Numerical calculations \cite{clement2005,modugno2006,akkermans2006}
confirm this scenario for parameters
relevant to the experiments of Refs.~\cite{clement2005,clement2006,fort2005}.

In this Letter, we show that the expansion of a 1D interacting BEC can exhibit 
AL in a random potential without large or wide modulations. Here, in contrast 
to the situation in Refs.~\cite{clement2005,clement2006,fort2005}, the BEC is not significantly 
affected by a single reflection.
For this {\it weak disorder} regime we have identified the following localization scenario 
on the basis of numerical calculations and the toy model described below.

At short times, the disorder does not play a significant role, atom-atom interactions drive the
expansion of the BEC and determine the long-time momentum distribution,
$\dist (k)$. 
According to the scaling theory \cite{scaling}, 
$\dist (k)$ has a high-momentum cut-off at $1/\xiini$, where
$\xiini=\hbar/\sqrt{4m\mu}$ and $\mu$ are the initial healing length and chemical potential 
of the BEC, and $m$ is the atom mass.
When the density is significantly decreased, the 
expansion is governed by the scattering of almost non-interacting
waves from the random potential. Each wave with momentum $k$
undergoes AL on a momentum-dependent length $\loc(k)$
and the BEC density profile will be determined by the superposition of 
localized waves.
For speckle potentials the 
Fourier transform of the correlation function 
vanishes for $k>2/\sigmar$, where $\sigmar$ is the correlation
length of the disorder, and the Born approach yields an 
effective mobility edge
at $1/\sigmar$.
Then, if the high-momentum cut-off is provided by the momentum 
distribution $\dist (k)$ (for $\xiini > \sigmar$), the BEC is
{\it exponentially} localized, whereas if the cut-off is provided by 
the correlation function of the disorder (for $\xiini < \sigmar$) the localization is {\it algebraic}.
These findings pave the way to observe AL in experiments
similar to those of Refs.~\cite{clement2005,clement2006,fort2005}.

We consider a 1D Bose gas with repulsive short-range
interactions, characterized by the 1D coupling constant $g$
and trapped in a harmonic potential $\Vh (z) = m\omega^2z^2/2$.
The finite size of the trapped sample provides 
a low-momentum cut-off for the phase fluctuations, and for
weak interactions ($n\gg mg/\hbar^2$, where $n$ 
is the 1D density), the gas forms a true BEC at low 
temperatures \cite{phasefluctoned}.

We treat the BEC wave function $\psi (z,t)$
using the Gross-Pitaevskii equation (GPE). In 
the presence of a superimposed random potential 
$\Vopt (z)$, this equation reads:
\begin{equation}
i\hbar \partial_t \psi
= \left[ \frac{-\hbar^2 }{2m}\partial_z^2 + \Vh (z) + \Vopt (z)
+ g|\psi|^2 - \mu \right] \psi, 
\label{eq:GPE} 
\end{equation}
where $\psi$ is normalized by $\int\! \textrm{d}z |\psi|^2 = N$, with $N$ being
the number of atoms. 
It can be assumed without loss of generality that the average of $V(z)$
over the disorder, $\langle V \rangle$, vanishes, while the correlation function
$C(z)=\langle V(z') V(z'+z)\rangle$ can be written as
$C(z) = \Vr^2 c(z/\sigmar)$,
where the reduced correlation function $c(u)$ has unity
height and width. 
So, $\Vr = \sqrt{\langle V^2 \rangle}$ 
is the standard deviation, and $\sigmar$ is
the correlation length of the disorder.

The properties of the correlation function depend on the model
of disorder.
Although most of our
discussion is general, 
we mainly refer to a 1D speckle random potential 
\cite{goodman} similar to the ones used in experiments with cold atoms 
\cite{specklegrynberg,lye2005,clement2005,clement2006,fort2005}.
It is a random potential with a 
truncated negative exponential single-point distribution
\cite{goodman}:
\begin{equation}
\mathcal{P}[V(z)] = 
\frac{\exp [ -(V(z)+\Vr)/\Vr ]}{\Vr}\
\Theta \left(\frac{V(z)}{\Vr} + 1\right),
\label{eq:speckle}
\end{equation}
where $\Theta$ is the Heaviside step function,
and with a correlation function which can be controlled almost at will
\cite{clement2006}.
For a speckle potential produced by diffraction through a 1D square 
aperture \cite{goodman,clement2006}, we have
\begin{equation}
C(z) = \Vr^2 c(z/\sigmar);\,\,\,\,\,\,\,\,c(u) = \sin^2 (u)/u^2.
\label{eq:specklecorr}
\end{equation}
Thus the Fourier transform of $C(z)$ has a finite support: 
\begin{equation}
\!\widehat{C}(k)\!=\!\Vr^2\sigmar \widehat{c}(k\sigmar);\, 
\widehat{c}(\kappa)\!=\!\sqrt{\pi/2}(1\!-\!\kappa/2)\Theta(1\!-\!\kappa/2),\!
\label{eq:corrfunc}
\end{equation}
so that $\widehat{C}(k)=0$ for $k>2/\sigmar$.
This is actually a general property of speckle potentials,
related to the way they are produced
using finite-size diffusive plates
\cite{goodman}.

We now consider the expansion 
of the BEC, using the following toy model.
Initially, the BEC is assumed to be at equilibrium in the
trapping potential $\Vh (z)$ and in the absence of disorder.
In the Thomas-Fermi 
regime (TF) where $\mu \gg \hbar\omega$, the initial BEC density
is an inverted parabola, $n(z)=(\mu/g)(1-z^2/\LTF^2)\Theta(1-|z|/\LTF)$,
with $\LTF=\sqrt{2\mu/m\omega^2}$ being the TF half-length. 
The expansion is induced
by abruptly switching off the confining trap at time $t=0$,
still in the absence of disorder. 
Assuming that the condition of weak interactions
is preserved during the expansion,
we work within the framework of the GPE~(\ref{eq:GPE}).
Repulsive atom-atom interactions drive the short-time ($t \lesssim 1/\omega$) 
expansion, 
while at longer times ($t \gg 1/\omega$) the interactions are not
important and the expansion becomes free. 
According to the scaling approach \cite{scaling}, 
the expanding BEC acquires a dynamical phase 
and the density profile is rescaled, remaining an inverted parabola:
\begin{equation}
\psi (z,t)=\left(\psi [z/b(t),0]/\sqrt{b(t)}\right)
\exp{\{\i m z^2 \dot{b}(t)/2\hbar b(t)\}}, 
\label{eq:scaling}
\end{equation}
where the scaling parameter $b(t)=1$ for $t=0$, and $b(t)\simeq \sqrt{2}\omega t$ 
for $t\gg 1/\omega$ \cite{clement2005}.

We assume that the random potential is abruptly switched on at a 
time $t_0 \gg 1/\omega$. Since the atom-atom interactions are no longer
important, the BEC represents a superposition of almost independent 
plane waves:
\begin{equation}
\psi(z,t) = \int\! \frac{\textrm{d}k}{\sqrt{2\pi}} \widehat{\psi} (k,t) \exp(\i kz).
\label{eq:TFwf}
\end{equation}
The momentum distribution $\dist (k)$ follows from Eq.~(\ref{eq:scaling}).
For $t \gg 1/\omega$, it is stationary and has a high-momentum cut-off
at the inverse healing length $1/\xiini$:
\begin{equation}
\dist (k) = |\widehat{\psi}(k,t)|^2 \simeq
\frac{3N\xiini}{4}(1-k^2\xiini^2)
\Theta(1-k\xiini),
\label{eq:scalingdist}
\end{equation}
with the normalization condition $\int_{-\infty}^{+\infty}\! \textrm{d}k \dist (k)=N$.

According to the Anderson theory \cite{anderson1958}, $k$-waves
will exponentially localize as a result of multiple scattering from the 
random potential. Thus, components
$\exp(\i kz)$ in Eq.~(\ref{eq:TFwf}) will become localized 
functions $\phi_k (z)$. At large distances, $\phi_k (z)$
decays exponentially, so that $\ln|\phi_k(z)|\simeq -\lyap (k) |z|$,
with $\lyap (k)=1/L (k)$ the Lyapunov exponent, and $\loc (k)$
the localization length.
The AL of the BEC occurs when the independent $k$-waves have localized.
Assuming that the phases of the functions $\phi_k (z)$, which are determined by
the local properties of the random potential and by the time $t_0$, 
are random, uncorrelated functions for different momenta, the BEC density is 
given by 
\begin{equation}
n_0(z)\equiv \langle | \psi (z) |^2 \rangle =
2\int_0^{\infty}\! \textrm{d}k \dist(k) \langle | \phi_k (z) |^2\rangle,
\label{eq:locBEC}
\end{equation}
where we have taken into account that $\dist (k)=\dist (-k)$ 
and $\langle | \phi_k (z) |^2\rangle = \langle | \phi_{-k} (z) |^2\rangle$.

We now briefly outline the properties of the functions $\phi_k(z)$
from the theory of localization of single particles.
For a weak random potential, using the phase formalism
\cite{lifshits1988} the state with momentum $k$ 
is written in the form:
\begin{equation}
\phi_k(z) = \phantom{k} r(z) \sin \left[\theta(z)\right];~
\partial_z \phi_k = k r(z) \cos \left[\theta(z)\right],
\label{eq:phasefunct}
\end{equation}
and the Lyapunov exponent is obtained from the relation
$\lyap (k) = - \lim_{|z| \to \infty}\langle \log \left[ r(z) \right] / |z| \rangle$.
If the disorder is sufficiently weak, then the phase is approximately $kz$ 
and solving the Schr\"odinger equation up to first order in
$|\partial_z \theta(z)/k - 1|$, one finds \cite{lifshits1988},
\begin{equation}
\lyap (k) \simeq 
(\sqrt{2\pi}/8\sigmar) (\Vr/E)^2 (k\sigmar)^2
\widehat{c} (2k\sigmar),
\label{eq:lyapunov2}
\end{equation}
where $E=\hbar^2k^2/2m$.
Such a perturbative (Born) approximation assumes the inequality
\begin{equation}
V_R\sigmar\ll (\hbar^2 k/m)(k\sigmar)^{1/2},
\label{eq:lyapunovcond}
\end{equation}
or equivalently $\lyap (k)\ll k$.
Typically, Eq.~(\ref{eq:lyapunovcond}) means that the random potential
does not comprise large or wide peaks.

Deviations from a pure exponential decay of $\phi_k$ are obtained 
using diagrammatic methods \cite{gogolin1976}, and one has
\begin{eqnarray}
\langle| \phi_k (z) |^2\rangle 
& = & \frac{\pi^2 \lyap (k)}{2} \int_0^\infty\! \textrm{d}u\ 
u\ \sh (\pi u) \times 
\label{eq:gogolin1} \\
&& \left( \frac{1+u^2}{1+\ch (\pi u)} \right)^2 
\textrm{exp}\{- 2 (1+u^2) \lyap (k) |z|\},
\nonumber
\end{eqnarray}
where $\lyap(k)$ is given by Eq.~(\ref{eq:lyapunov2}).
Note that at large distances ($\lyap (k)|z| \gg 1$), Eq.~(\ref{eq:gogolin1})
reduces to
$\langle| \phi_k (z) |^2\rangle 
\simeq  \Big(\pi^{7/2}/64\sqrt{2\lyap (k)}|z|^{3/2}\Big) 
\exp\{-2\lyap (k)|z|\}$.

The localization
effect is closely related to the properties of  the correlation function of the 
disorder.
For the 1D speckle potential the correlation function 
$\widehat{C} (k)$ 
has a high-momentum cut-off $2/\sigmar$, and from Eqs.~(\ref{eq:corrfunc})
and (\ref{eq:lyapunov2}) we find
\begin{equation}
\!\lyap(k)\!=\!\lyap_0(k)(1\!-\!k\sigmar)\Theta(1\!-\!k\sigmar);\,
\lyap_0(k)\!=\!\frac{\pi m^2\Vr^2\sigmar}{2\hbar^4 k^2}.
\label{eq:lyapunovspeckle}
\end{equation}
Thus, one has
$\lyap (k)>0$ only for $k\sigmar < 1$ 
so that there is a mobility edge at $1/\sigmar$ in the Born approximation.
Strictly speaking, on the basis of this approach one cannot say that the
Lyapunov exponent is exactly zero for $k>1/\sigmar$.
However, direct numerical calculations of the Lyapunov exponent show that
for $k>1/\sigmar$ it is at least two orders of magnitude smaller than
$\lyap_0(1/\sigmar)$ representing a characteristic value of $\lyap (k)$
for $k$ approaching $1/\sigmar$. 
For $\sigmar \gtrsim 1\mu m$, achievable for speckle potentials \cite{clement2006} and
for $\Vr$ satisfying Eq.~(\ref{eq:lyapunovcond}) with $k \sim 1/\sigmar$,
the localization length at $k>1/\sigmar$ exceeds
$10$cm which is much larger than the system size in the studies of
quantum gases.
Therefore, $k=1/\sigmar$ corresponds to an effective mobility edge
in the present context.
We stress that it is a {\it general} feature 
of optical speckle potentials, owing to the finite support of the
Fourier transform of their correlation function.

We then use Eqs.~(\ref{eq:scalingdist}), (\ref{eq:gogolin1}) and (\ref{eq:lyapunovspeckle}) 
for calculating the density profile of the localized 
BEC from Eq.~(\ref{eq:locBEC}). Since the high-momentum cut-off of
$\dist (k)$ is $1/\xiini$, 
and for
the speckle potential the cut-off of $\lyap(k)$ is $1/\sigmar$, the upper bound of integration 
in Eq.~(\ref{eq:locBEC}) is $\kc={\rm min}\{1/\xiini,\,1/\sigmar\}$.  
As the density profile $n_0(z)$ is a sum of functions
$\langle | \phi_k (z) |^2 \rangle$ which decay exponentially
with a rate $2\lyap (k)$, the long-tail behavior of $n_0(z)$ 
is mainly determined by the components with the smallest $\lyap (k)$,
\ie those with $k$ close to $\kc$, 
and integrating in Eq.~(\ref{eq:locBEC}) we limit ourselves to leading order terms
in Taylor series for $\dist (k)$ and $\lyap (k)$ at $k$
close to $\kc$.

For $\xiini>\sigmar$, the high-momentum cut-off $\kc$ in Eq.~(\ref{eq:locBEC}) is
set by the momentum distribution $\dist (k)$ and is equal to $1/\xiini$. 
In this case all functions $\langle | \phi_k (z) |^2 \rangle$ have a finite 
Lyapunov exponent, $\lyap (k) > \lyap (1/\xiini)$, and
the whole BEC wave function is {\it exponentially localized}. For the long-tail 
behavior of $n_0(z)$, from Eqs.~(\ref{eq:scalingdist}), (\ref{eq:locBEC}) and 
(\ref{eq:gogolin1}) we obtain: 
\begin{equation}
n_0(z)\propto |z|^{-7/2}\exp\{-2\lyap(1/\xiini)|z|\};\,\,\,\,\xiini>\sigmar.
\label{eq:asympt1}
\end{equation}
Equation (\ref{eq:asympt1}) assumes the inequality $\lyap(1/\xiini)|z|\gg 1$, 
or equivalently $\lyap_0(\kc)(1-\sigmar/\xiini)|z|\gg 1$.

For $\xiini<\sigmar$, $\kc$ is provided by 
the Lyapunov exponents of $\langle |\phi_k(z)|^2\rangle$ so that they do not have a finite lower bound. Then
the localization of the BEC becomes {\it algebraic} and it is only {\it partial}. The part
of the BEC wave function, corresponding to the waves with momenta in the range $1/\sigmar<k<1/\xiini$,
continues to expand. 
Under the condition
$\lyap_0(\kc)(1-\xiini^2/\sigmar^2)|z| \gg 1$ for the asymptotic
density  distribution of localized particles, 
Eqs.~(\ref{eq:locBEC}) and (\ref{eq:gogolin1}) yield:  
\begin{equation}
n_0(z)\propto |z|^{-2};\,\,\,\,\xiini<\sigmar.
\label{eq:asympt2}
\end{equation}

Far tails of $n_0(z)$ will be always described by the asymptotic relations 
(\ref{eq:asympt1}) or (\ref{eq:asympt2}), unless $\xiini=\sigmar$. In the special case of 
$\xiini=\sigmar$, or for $\xiini$ very close to $\sigmar$ and at distances where
$\lyap_0(\kc)|(1-\xiini^2/\sigmar^2)z| \ll 1$,
still assuming that $\lyap_0(\kc)|z|\gg 1$ we find  
$n_0(z)\propto |z|^{-3}$.

Since the typical momentum of the expanding BEC is
$1/\xiini$, according to Eq.~(\ref{eq:lyapunovcond}), our approach 
is valid for $\Vr\ll \mu(\xiini/\sigmar)^{1/2}$. For a speckle
potential, the typical momentum of the waves which become localized is $1/\sigmar$
and for $\xiini<\sigmar$ the restriction is stronger: $V_R\ll\mu(\xiini/\sigmar)^2$. 
These conditions were not fulfilled, neither in the experiments of
Refs.~\cite{clement2005,clement2006,fort2005}, nor in the numerics
of Refs.~\cite{clement2005,modugno2006,akkermans2006}.

We now present numerical results for the expansion
of a 1D interacting BEC in a speckle potential, performed on the
basis of Eq.~(\ref{eq:GPE}). The BEC is initially at equilibrium 
in the combined random plus harmonic potential,
and the expansion of the BEC is induced by switching off abruptly
the confining potential at time $t=0$ as in
Refs.~\cite{fort2005,clement2005,clement2006,lsp2005}.
The differences from the model discussed above are that the random potential 
is already present for the initial stationary condensate and 
that the interactions are maintained during the whole expansion.
This, however, does not significantly change the physical picture.

\begin{figure}[b!]
\begin{center}
\infig{28.em}{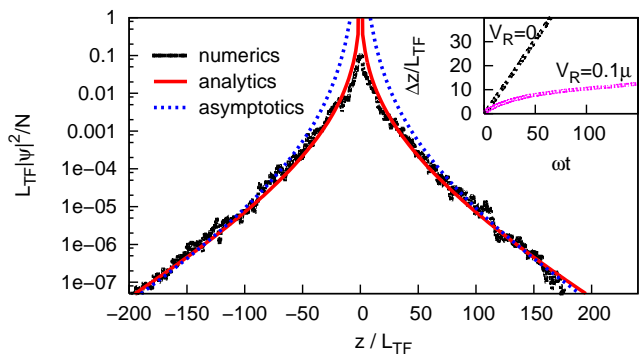}
\end{center}
\caption{(color online)
Density profile of the localized BEC in a
speckle potential at $t=150/\omega$. Shown are 
the numerical data (black points),
the fit of the result from Eqs.~(\ref{eq:scalingdist}), (\ref{eq:locBEC}) 
and (\ref{eq:gogolin1}) [red solid line], and the fit of 
the asymptotic formula~(\ref{eq:asympt1}) [blue dotted line].
Inset: Time evolution of the rms size of the BEC.
The parameters are $\Vr=0.1\mu$, $\xiini=0.01 \LTF$,
and $\sigmar=0.78\xiini$.
} 
\label{fig:density0.55Vr0.1}
\end{figure}
 
The properties of the initially trapped BEC have been discussed 
in Ref.~\cite{lsp2006} 
for an arbitrary ratio $\xiini/\sigmar$.
For $\xiini \ll \sigmar$, the BEC follows the
modulations of the random potential, while for
$\xiini\gtrsim\sigmar$ the effect of the random potential can be significantly smoothed. 
In both cases, the weak 
random potential only slightly modifies the density profile \cite{lsp2006}. 
At the same time, the expansion of 
the BEC is strongly suppressed compared to the non-disordered case. 
This is seen from the time evolution of the rms size of the BEC,
$\Delta z = \sqrt{\langle z^2 \rangle - \langle z \rangle^2}$,
in the inset of Fig.~\ref{fig:density0.55Vr0.1}. 
At large times, the BEC density reaches an almost stationary profile.
The numerically obtained density profile in 
Fig.~\ref{fig:density0.55Vr0.1} shows an excellent agreement 
with a fit of $n_0(z)$ from Eqs.~(\ref{eq:scalingdist}), (\ref{eq:locBEC}) and
(\ref{eq:gogolin1}), where a multiplying constant was the only fitting parameter. 
Note that Eq.~(\ref{eq:locBEC}) overestimates the density in the
center of the localized BEC, where the contribution of waves with very small $k$
is important. This is because Eq.~(\ref{eq:lyapunovspeckle}) overestimates $\lyap (k)$ 
in this momentum range, where the criterion (\ref{eq:lyapunovcond}) is not satisfied.

\begin{figure}[t!]
\begin{center}
\infig{28.em}{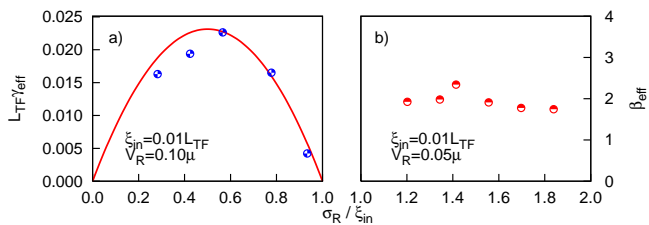}
\end{center}
\caption{(color online)
a) Lyapunov exponent $\lyap_\textrm{eff}$ in units of $1/\LTF$ 
for the localized BEC in a speckle potential, in the regime $\xiini>\sigmar$.
The solid line is $\lyap(1/\xiini)$ from Eq.~(\ref{eq:lyapunovspeckle}).
b) Exponent of the power-law decay of the localized BEC in the regime 
$\xiini<\sigmar$. The parameters are indicated in the figure.
} 
\label{fig:lyap}
\end{figure}

We have also studied the long-tail asymptotic behavior of the numerical data. 
For $\xiini>\sigmar$, we have performed fits of
$|z|^{-7/2} \textrm{e}^{-2\lyap_\textrm{eff}|z|}$ to the data.
The obtained $\lyap_\textrm{eff}$
are in excellent agreement with $\lyap(1/\xiini)$ following from
the prediction of Eq.~(\ref{eq:asympt1}), as shown in Fig.~\ref{fig:lyap}a.
For $\xiini<\sigmar$, we have fitted 
$|z|^{-\beta_\textrm{eff}}$ to the data.
The results are plotted in Fig.~\ref{fig:lyap}b and show that the long-tail
behavior of the BEC density is compatible with a power-law
decay with $\beta_\textrm{eff} \simeq 2$, in agreement with the 
prediction of Eq.~(\ref{eq:asympt2}).

In summary, we have shown that in weak disorder the expansion
of an initially confined interacting 1D BEC can exhibit Anderson localization. 
Importantly, the high-momentum cut-off of the Fourier transform
of the correlation function for 1D speckle potentials can change 
localization from exponential to algebraic. 
Our results draw prospects for the observation of Anderson localization
of matter waves in experiments similar to those of 
Refs.~\cite{clement2005,fort2005,clement2006}.
For $\Vr=0.2\mu$, $\xiini=3\sigmar/2$ and $\sigmar=0.27\mu$m,
we find the localization length $L(1/\xiini)\simeq 460\mu$m.
These parameters are in the range of accessibility of current experiments 
\cite{clement2006}. In addition,
the localized density
profile can be imaged directly, which allows one to distinguish between
exponential and algebraic localization. 
Finally, we would like to raise 
an interesting problem for future studies.
The expanding and then localized BEC is an excited Bose-condensed state
as it has been made by switching off the confining trap.
Therefore, the remaining small interaction between atoms 
should cause the depletion of the BEC and the relaxation to a new equilibrium
state. The question is how the relaxation process occurs and to which extent it 
modifies the localized state. 

We thank M.~Lewenstein, S.~Matveenko, P.~Chavel, P.~Leboeuf and N.~Pavloff for 
useful discussions. 
This work was supported by 
the French DGA, IFRAF,
Minist\`ere de la Recherche (ACI Nanoscience 201),
ANR (grants NTOR-4-42586, NT05-2-42103 and 05-Nano-008-02),
and
the European Union (FINAQS consortium and grants IST-2001-38863 and MRTN-CT-2003-505032),
the ESF program QUDEDIS, 
and the Dutch Foundation FOM.
LPTMS is a mixed research unit 8626 of CNRS and University Paris-Sud.


\end{document}